# Spherulitic and rotational crystal growth of Quartz thin films

Nick R. Lutjes, Silang Zhou, Jordi Antoja-Lleonart, Beatriz Noheda, Václav Ocelík[*]

*Zernike Institute for Advanced Materials, University of Groningen, Nijenborgh 4, 9747 AG Groningen, The Netherlands*

**Abstract**

To obtain crystalline thin films of alpha-Quartz represents a challenge due to the tendency for the material towards spherulitic growth. Thus, understanding the mechanisms that give rise to spherulitic growth can help regulate the growth process. Here the spherulitic type of 2D crystal growth in thin amorphous Quartz films was analyzed by electron back-scatter diffraction (EBSD). EBSD was used to measure the size, orientation, and rotation of crystallographic grains in polycrystalline $SiO_2$ and $GeO_2$ thin films with high spatial resolution. Individual spherulitic Quartz crystal colonies contain primary and secondary single crystal fibers, which grow radially from the colony center towards its edge, and fill a near circular crystalline area completely. During their growth, individual fibers form so-called rotational crystals, when some lattice planes are continuously bent. The directions of the lattice rotation axes in the fibers were determined by an enhanced analysis of EBSD data. A possible mechanism, including the generation of the particular type of dislocation(s), is suggested.

---

[*] Corresponding author: v.ocelik@rug.nl    tel: +31 50 363 3407    fax: +31 50 363 4441

# Introduction

The synthesis of piezoelectric epitaxial Quartz thin films with submicron thicknesses seems to be a serious challenge towards further miniaturization of oscillator circuits. Recently, we reported the successful solid-state crystallization of pure $GeO_2$ amorphous thin films into a α-Quartz (piezoelectric) structure via control of the crystallization temperature[1]. Pulsed laser deposition (PLD) was used to prepare approximately 140 nm-thick amorphous films of $GeO_2$ on different crystalline substrates. It has been observed, that columnar and dendritic growth take place at high crystallization temperatures (> 870 ºC). Spherulitic crystal growth was promoted at lower temperatures (< 850 ºC), often accompanied by a gradual crystal rotation inside individual spherulite fibers. Two-dimensional spherulites are made of fibers that grow radially, starting from one nucleation core and conferring the typical circular shape. Generally, spherulites are formed by a non-crystallographic branching mechanism[2–4], where the parent and daughter fibers do not share the same crystal orientation. The misorientation angles between neighbor fibers typically range between 0-15°[3]. The non-crystallographic branching mechanism of growth distinguishes spherulites from other-branched crystals and polycrystalline aggregates possessing round forms[3].

A 2D spherulitic growth during solid recrystallization of an amorphous thin film is not new. Pure selenium, Se containing compounds, as well as α-$Fe_2O_3$, and some organic chemistry compounds showing spherulitic crystallization have been reviewed[3]. However, the unique 2D spherulitic growth Quartz structure has not been reported thus far. Some basics of trigonal symmetry of α-Quartz crystal are summarized in Supplementary Information (SI), together with a short notation of few important crystallographic planes[5].

Scanning electron microscopy (SEM) and Atomic Force Microscopy (AFM), which are typically used to identify growth morphologies, often fail to differentiate between spherulites and crystallographic grains, due to the fact they are unable to provide crystallographic information[6]. Instead, Electron Back-Scatter Diffraction (EBSD), which offers a combination of the high lateral resolution of SEM and information about local crystal orientation, is an excellent technique to study these objects.

Besides non-crystallographic branching leading to the formation of individual fibers in Quartz thin film crystallization[1], a gradual crystal lattice rotation inside each fiber is also observed through EBSD. This kind of crystal growth has already been reported as micron-sized trans-rotational 2D crystals of FeO[7] and SbTe[8], formed by thermally-induced crystallization of amorphous thin films during in-situ experiments in a transmission electron microscope. These films were not thicker than 40 nm. However, lattice rotation gradients were sometimes extremely high (up to 100 °/µm). Savytskii and co-authors[9] reported large scale rotating lattice single crystals growth on the surface of bulk $Sb_2S_3$ glass. Surface crystallization was the result of a solid-state transformation, induced by local surface heating with a continuous-wave laser beam spot. Subsequently, the laser spot movement on the sample surface with a constant speed resulted in a crystalline track of about 1.5 µm thick, 5 µm wide and up to 90 µm long. Different crystal rotation gradients between 0 and 0.6 °/µm were observed. The authors in that study also used EBSD to determine crystal lattice orientation, combined with scanning X-ray micro-diffraction, which has much higher orientational accuracy. EBSD has relatively good spatial resolution (in our case estimated as 50 nm). However, its accuracy in the determination of crystal orientation is considered to be relatively low (typically 1-2°), and dependent on many experimental parameters. Nonetheless, the angular resolution in the determination of misorientation between two points on the same EBSD map is one order of magnitude better[10]. This is an important fact, because in our further analysis we will also determine a second misorientation parameter, i.e. the misorientation axis. This helps to study the origin and character of the crystal strain, and to understand the role of dislocations in such rotational crystal growth.

Changes in crystal orientation during its plastic deformation occur by the action of dislocations on specific crystallographic slip systems. Each slip system comprises the slip plane and the slip direction within this plane. Importantly, the deformation by a particular slip system is associated with a crystal rotation along an axis characteristic for that slip system[11]. This rotation axis is orthogonal simultaneously to the slip plane normal and the slip direction. The Inverse Pole Figure (IPF), given in Fig. S1c) in the SI, shows directions of the crystal rotation axes characteristic for all experimentally observed dislocation slip systems in Quartz. Details of these slip systems are summarized in Table S1 in SI.

Thus the scientific goal of our study is a detailed description of the 2D spherulitic crystallization of $SiO_2$ and $GeO_2$ amorphous thin films using EBSD. Analysis of lattice misorientation data should also clarify the origin and mechanisms of Quartz crystal deformations during trans-rotational growth of individual spherulitic fibers.

## Experimental

Thin films of $GeO_2$ were deposited on $MgAl_2O_4$ substrates (CrysTec GmbH) with a 248 nm KrF pulsed laser (Lambda Physik COMPex Pro 205). The Pulsed Laser Deposition (PLD) parameters were reported in detail elsewhere [1]. After deposition, the amorphous film was annealed at 830 °C for 30 minutes. The thicknesses of the thin films were estimated to be about 120 nm from X-ray reflectivity (Panalytical X'Pert, CuKα radiation) and the surface morphology was imaged by an AFM (Bruker Dimension Icon).

To create $SiO_2$ amorphous thin films, we deposit Sr particles on the p-Si(100) substrates by ablating a target of compacted, dendritic Sr in the PLD chamber using 30 pulses at 1 Hz repetition rate. Afterwards, the sample was transported in air to a Picosun R-200 Advanced hot-wall Atomic Layer Deposition (ALD) system. Using bis(diethylamino)silane (BDEAS) and ozone (generated by an INUSA Ozone Generator) as precursors, a $SiO_2$ layer was grown, at a process temperature of 300 ºC. We use 600 ALD cycles, which on a blank Si substrate grow 25 nm of $SiO_2$ using the same process parameters. Finally, the sample was annealed by ramping at 20 ºC/min to 1000 ºC, staying there for 5 h, and then cooling down passively to room temperature, all under an oxygen flow of 12 L/h.

The annealed films were studied using a FEI Nova NanoSEM field emission gun scanning electron microscope (SEM), equipped with Electron Back-Scatter Diffraction (EBSD) from EDAX. An acceleration voltage of 20 keV and a beam current of about 2 nA were used, together with the low vacuum mode (chamber pressure of 50 Pa) to avoid sample surface charging[12]. The sample normal was tilted 71º towards the electron beam.

Kikuchi patterns were framed using a Hikari CCD camera with speed of 10–30 frames/s. EBSD data were collected by OIM DC 7.3 software, using online pattern indexing[13] by detection of 12-14 Kikuchi bands to avoid "pseudo-symmetry" problems reported for Quartz-type crystals[14]. It has been experimentally determined that the above mentioned parameters result in a successful collection of EBSD data when the step size between neighboring scanning points (hexagonal grid) was not smaller than 100 nm. Smaller step size, higher electron beam acceleration voltage, larger electron beam current or slower map scanning speed usually resulted in a local crystal damage, which limits the resolution of the scan in order to retain good quality Kikuchi patterns for the complete beam scanning area.

OIM Analysis 8.1 software together with MTEX[15], a free extension for Matlab®, were used to: i) clean and analyze EBSD data; ii) calculate grain boundaries and crystal misorientations; iii) plot EBSD maps, pole figures and orientation distributions. A gentle, two step data cleaning procedure was applied to remove speckle points from EBSD maps. The procedure starts with Grain Confidence Index Standardization (setting a grain tolerance angle of 3º, minimum grain size of 5 pixels at multiple rows), followed by a Neighbor Orientation

Correlation procedure (level 4, tolerance 3, and a minimal Confidence Index (CI) of 0.1). Only the second data cleaning treatment step may change the crystal orientation in some scanning points (typically < 4%). Finally, we remove all data points where the CI remains smaller than 0.1. In these points the determined crystal orientation is not reliable, and therefore these points were dismissed before performing any calculations and plots (shown as white points in the presented EBSD maps). Trigonal Quartz crystal symmetry is properly colorized in crystal orientation maps generated by MTEX package[16], and therefore the maps are plotted via this tool in this work.

## Results and discussion

Typical results of EBSD observation of a $GeO_2$ thin film after crystallization heating are shown in Fig. 1. The film exhibits relatively large crystalline areas (from tens to hundreds of μm in size) with Quartz crystal symmetry, as the IPF map (crystal direction normal to the sample surface) in Fig. 1a) clearly demonstrates. Crystalline areas are sometimes isolated by an amorphous surrounding, occasionally touching each other and forming features resembling incomplete 2D Voronoi polyhedral tessellation[17]. These are formed via: i) nucleation in randomly distributed points; followed by ii) stellular 2D growth from nucleation centers, sometimes isolated in an amorphous surrounding, sometimes halted in their growth by the presence of another crystalline object that grew from a neighboring nucleation point. The crystalline objects, shown in Fig. 1a), clearly exhibit all characteristics of spherulitic crystal growth reviewed by Shtukenberg et al.[3].

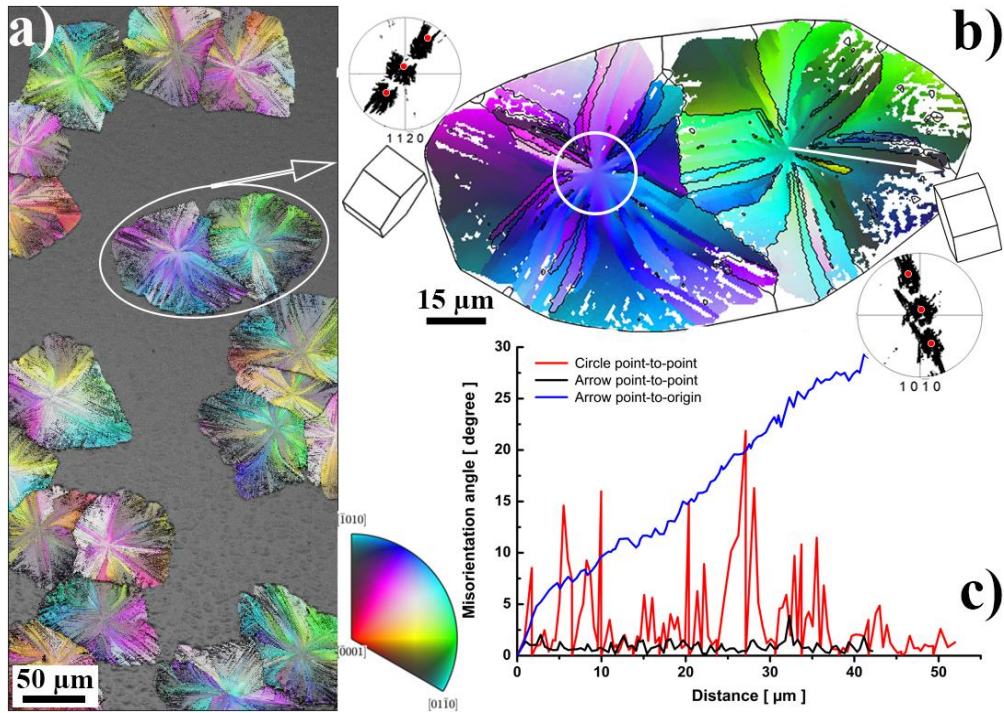

*Fig. 1: EBSD observations of spherulitic GeO$_2$ Quartz crystal islands.*
*a) IPF map (color) of crystal directions perpendicular to the thin film surface, combined with Image Quality map (gray scale);*
*b) The same IPF map of two selected spherulites with grain boundaries (> 3º) marked by black lines. Side insets show the axonometric projections of crystal orientation at the center of these two spherulites, together with their corresponding ($11\bar{2}0$) and ($10\bar{1}0$) pole figures, on which orientations of the spherulite centers are marked by red points;*
*c) Crystal misorientation angle profiles measured: i) along the white circle around the center of the left spherulite on b); ii) along the white arrow from the center to the edge of one fiber of the right spherulite on b).*
*The crystal direction color key is valid for both maps shown in a) and b).*

A detailed IPF map shown in Fig. 1b) displays grain boundaries between individual crystallites having mutual misorientations larger than 3º. Two inserted axonometric projections of the Quartz crystal lattice depict the crystal lattice orientation of the central part of both spherulites in Fig. 1b). The spherulite on the left side nucleates with its <$11\bar{2}0$> direction almost parallel to the film surface normal. In the case of the right spherulite, nucleation happened with the crystal direction <$10\bar{1}0$> almost parallel to the surface film normal. These two nucleation orientations are also marked in the attached ($11\bar{2}0$) and ($10\bar{1}0$) pole figure insets as red points. These two pole figures contain the orientations of all corresponding spherulite points, forming clouds of small black dots. It is clear that the nucleus orientation defines "a main" orientation and all other crystal orientations in the spherulite slightly deviate from it as the spherulitic crystalline object grows. This growth is realized either via primary fiber growth from the nucleation center without the formation of small angle grain boundaries (> 3º), or via so-called non-crystallographic branching[3], when new sub-crystals (secondary fibers) with a small misorientation (3 - 20º) heterogeneously nucleate on the side of an already growing fiber and grow further as new fibers, also mostly in radial directions.

Internal and inter-fiber crystal misorientations are shown in Fig. 1c). The point-to-point crystal misorientation angle along the white circle around the nucleation point of the left spherulite (Fig. 1b), is shown by the red line. The point-to-point misorientation angle inside one fiber does not exceed 3º, while misorientations at fiber boundaries vary between 3º and 22º, which is typical for spherulitic growth[3]. Non-crystallographic branching is probably a consequence of anisotropic crystal growth. Main primary fibers grew radially from the nucleation center in directions of fast Quartz crystal growth, leaving some amorphous residual areas in between. These areas then crystallize via nucleation and subsequent growth of secondary or even tertiary fibers. All these "daughter" fibers have their orientations slightly different from primary ones (3 - 25º), with preferable grow directions being able to fill amorphous residuals between primary fibers. A slight misorientation between primary and secondary fibers is the consequence of the aforementioned non-crystallographic branching, typical for spherulitic crystallization[3].

However, as further analysis has shown, Quartz crystal fibers do not grow as a crystal with a single, fixed crystallographic orientation. For instance, the black and the blue lines in Fig. 1c), respectively show the point-to-point and point-to-origin crystal lattice misorientation profile along the white arrow running in one primary fiber, from the center to the edge of the spherulite on the right side of Fig. 1b). The fiber growth is, thus, clearly associated with a small misorientation between neighboring fiber points, that gradually accumulates and gives rise to a larger overall misorientation between the nucleation point and all other points of the fiber. The misorientation angle increases almost linearly with the distance from the nucleation center, achieving relatively large values at the edge of the spherulite. The average gradient of crystal rotation angle, $grad(\varphi)$ along the axis of the primary fiber selected in Fig. 1b) was estimated as 0.68°/µm, and the overall accumulated misorientation between the nucleation center and the edge of the spherulite amounts to 29°. Such behavior of monotonous increase of crystal misorientation angle along the fiber growth direction was observed for all randomly selected fibers inside all spherulites presented in Fig. 1a, with gradient variations between 0.3 – 1.5 °/µm. The monotonous increase character of the blue curve in Fig. 2c) suggests that, during growth, the Quartz crystal gradually rotates over the same axis, with a small rotation angle between neighboring points, but a much larger accumulated rotation angle for a longer distance. The above mentioned crystal rotation gradients correspond to crystal bending with curvatures between $5 \cdot 10^3$ and $26 \cdot 10^3$ m$^{-3}$.

During the collection of EBSD data, crystal orientation is measured at every point of the scanning area and expressed as a triplet of Euler angles, determining the crystal orientation with respect to the Cartesian sample coordinate system[18]. However, to study misorientation between two different crystal orientations, when one of them is chosen to be the reference orientation, the so-called angle/axis misorientation formulation is often used. In this approach, the original Euler angles data of both orientations are used to find the crystal direction that is not changed, when the crystal with the reference orientation is transformed to another one. The whole transformation can then be described by a rotation around an axis having such a crystal direction, using the rotation angle that transforms one crystal orientation to another[19].

Figure 2 summarizes the results of a detailed crystal misorientation analysis of EBSD data of one single primary fiber as studied in Fig. 1c), analyzed using the following procedure. The point with crystal orientation closest to the spherulite's nucleation orientation has been marked as the "reference point". The misorientation (containing the misorientation angle and rotation axis) between this reference point and all other points in the fiber has been calculated.

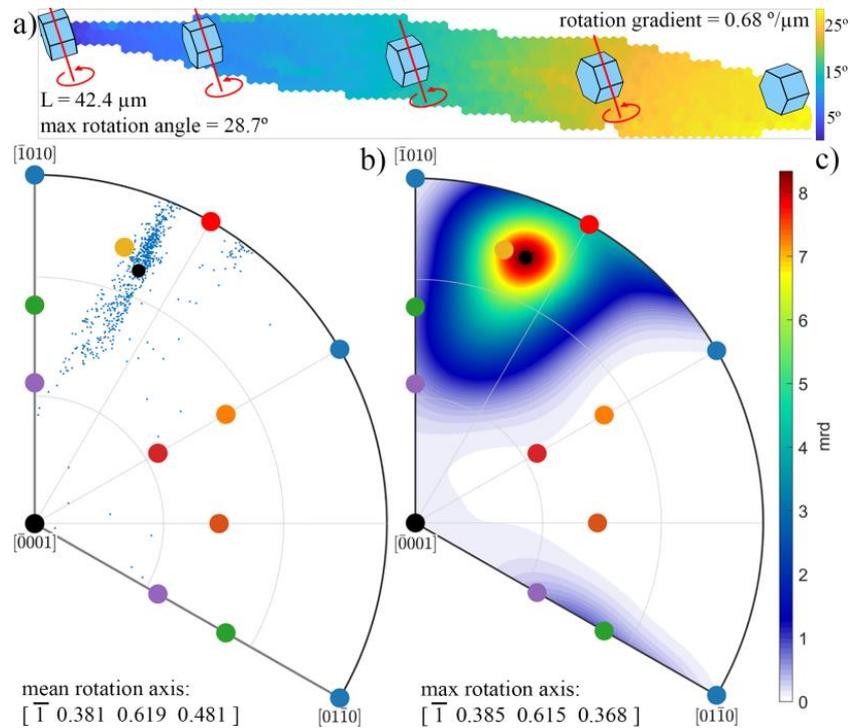

Fig. 2: Crystal misorientation analysis of an individual fiber;
  a) Misorientation angle map of one horizontal fiber from the right spherulite on Fig. 1b. Color denotes the misorientation angle, between a local point and the leftmost fiber point (close to spherulite nucleation orientation). Local lattice orientation is shown in a few points, together with rotation axis and direction;
  b) IPF of the rotation axis directions observed between the most left point and all other points in the fiber. The mean rotation axis direction is shown by a small black circle;
  c) IPF of the calculated rotation axis distribution. The distribution maximum is shown by a small black circle;
  Rotation axes of all common slip systems from Fig. S1c) and Table S1 from SI are also marked in both IPFs.

Figure 2a) shows the misorientation angle size map of the whole fiber, with clear evidence that the accumulated misorientation angle increases with distance. Axonometric projections of the Quartz crystal orientation demonstrate the local lattice orientation in a few selected points of the whole fiber. From the maximum misorientation angle (28.7°) and the whole fiber length (42.4 µm), the lattice rotation gradient is estimated to be 0.68 °/µm.

The rotation axis directions calculated for all 754 fiber points are shown via IPF in Fig. 2b). Few rotation axes seem to be distributed randomly, which corresponds with misorientations calculated for fiber points located close to the reference pixel. For these points, the measured value of the misorientation angle and the determination of the misorientation axis is close to the resolution limit of EBSD technique[20], and therefore suffers from large experimental errors. In any case, the majority of rotation axis directions, shown in Fig. 2b), are well concentrated in the vicinity of one crystal direction marked by the yellow dot. In reality, the calculated mean rotation axis, as well as the maximum of their distribution (Fig. 2c)), are both very close to this rotation axis direction. This rotation axis belongs to the action of the {z}<c+a> slip system in Quartz (see Fig. S1c) in SI). The projection of this crystal direction is also shown on the misorientation map (Fig. 2a)) as red lines, together with an arrow showing the direction of rotation.

A short discussion concerning the elongated spread of misorientation directions, shown in IPF on Fig. 2b), is required. A possible cause of the elongated spread could be explained by the fact that we did not analyze misorientations between neighboring points. Such analysis should be correct, because the lattice strain is gradually built from one point to another by a continuous lattice rotation. However, the misorientation between neighboring points is too small for a proper rotation axis determination based solely on EBSD data. Therefore, we assume the same or very similar direction of misorientation axis between all neighboring points inside one fiber. Such assumption is supported by the monotonous increase of the misorientation angle with distance between points, and allows the analysis based on calculation of misorientations between fiber points at larger distances.

Savytskii and co-authors[9] determined the direction of the crystal rotation axis by observing a color change on IPF maps, plotted in three perpendicular sample directions. They concluded that the crystal lattice appears to rotate about the tangent axis that lies in the plane of the substrate and perpendicular to growth direction[21]. However, our analysis allows for a more precise determination of the mean direction of crystal lattice rotation during the fiber growth. We calculated a 94º angle between the averaged direction of the crystal rotation axis and the direction of fiber growth (see Fig. 2a). For all other analyzed fibers this angle is found to be close to 90º.

Based on this observation, we may conclude that the relatively simple model proposed by Kooi and De Hosson[8], is a good description of the observation. In this model, a crystal lattice plane, originally parallel to the surface, is gradually pushed and curved during rotational crystal growth in the direction towards the amorphous substrate, as a consequence of the decrease in specific volume over the crystallization front. The model assumes that the position of the crystallization front near the surface is ahead of this front near the substrate. In the model proposed by Savitskii et al.[9], the same assumption concerning the shape of the crystallization front is required. A generation of one type of dislocations, directly at the glass/crystal interface, is anticipated, again as a consequence of the different densities between the glassy and the crystalline state. However, even though both models use the same assumptions, the latter directly introduces crystal dislocations as carriers of crystal plasticity, while the former also allows pure elastic bending. In our analysis we will focus on the dislocation model[9], and use the assumption that all these dislocations generated at the glass/crystal interface form a cloud of geometrically necessary dislocations (GNDs) responsible for the "bending" of the whole crystal lattice inside the crystal. A substantial advantage of our analysis using current EBSD data is that we may determine two quantities simultaneously: i) the bending lattice strain from the gradient of the misorientation angle; and ii) the type of dislocations responsible for such bending, derived from the mean direction of the rotation axes, which according to Fig. S1c) and Table S1 from SI may identify responsible slip systems.

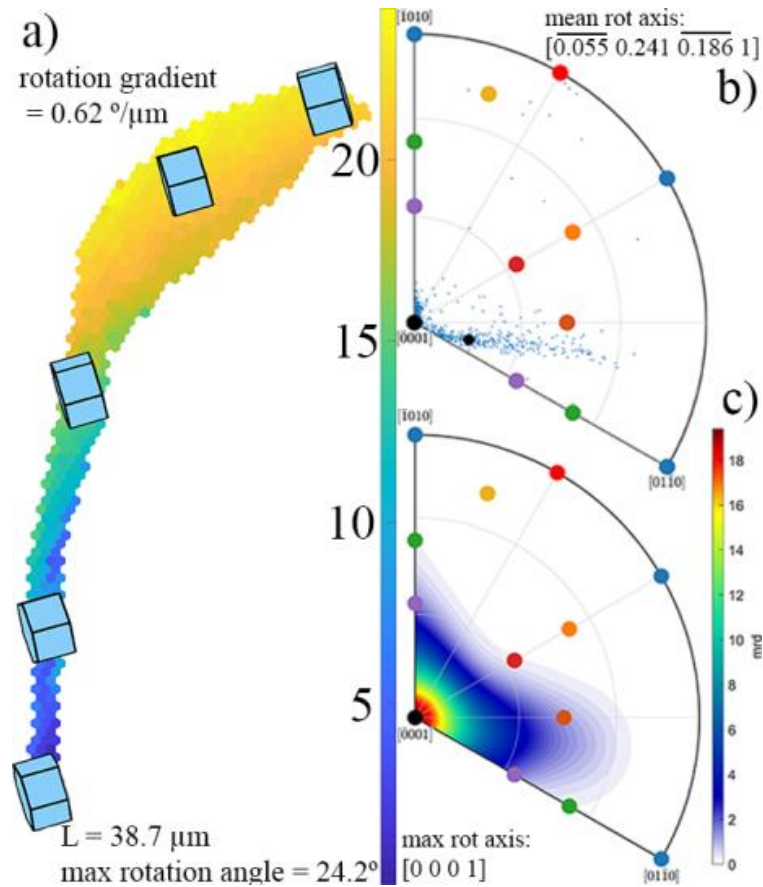

*Fig. 3: Misorientation analysis in an individual fiber;*
*a) Misorientation angle map of one vertical fiber in the right spherulite on Fig. 1b. Color denotes the misorientation angle, between a local point and the lowest fiber point (close to spherulite nucleation orientation). The local Quartz lattice orientation is shown in a few points.*
*b) Rotation axis directions observed between the lowest point and all other points in the fiber. The mean rotation axis direction is shown by a small black circle.*
*c) Calculated rotation axis distribution. The distribution maximum is located in the [0001] direction.*
*Rotation axes of all common slip systems from Fig. S1c) and Table S1 from SI are also marked in both IPFs.*

With the aim to prove this concept, a similar analysis has been performed on another fiber of spherulite shown on the right side of Fig. 1b). The fiber in Fig. 3a) grows from the same spherulite center as the fiber analyzed in Fig. 2. It initially grows in the north direction in the image, and gradually turns onto a north-east direction. The misorientation analysis is summarized in Fig. 3, which shows the misorientation map between the reference point and all other points of this fiber (see Fig. 3a) and the rotation axis direction analysis (see Fig. 3b-c).

Again, a clear increase of the misorientation angle with distance from the spherulite center is observed, as shown in Fig. 3a). A similar value of the gradient of lattice rotation angle (0.62 °/µm) was also detected in this fiber. However, as expected from the aforementioned model, the direction of the crystal lattice rotation axis is completely different. The reference point (the lowest point on the map shown in Fig. 3a)) has the same crystal orientation as the reference point in previous fiber analysis (Fig. 2), but the fiber initially grew in a different direction, about 90° apart from that of the fiber in Fig. 2. Using the assumption of the model, that is, a plane originally parallel to the sample surface (a crystal plane close to $(10\bar{1}0)$ in Fig. 3a), is being pushed towards the substrate during the fiber growth, one expects the lattice rotation axis direction to be close to [0001] crystal direction. The calculated mean and maximum rotation axis directions shown in Fig. 3b)-c) agree with this assumption: the maximum of the distribution of rotation axes is located exactly in the [0001] crystal direction, while the mean direction is just slightly inclined from it, along the $[\overline{0.055}\ 0.241\ \overline{0.186}\ 1]$ direction. Such a difference between the mean and the maximum direction is the result of an asymmetry in the distribution of the calculated rotation axis directions, between the reference point and all other points of the fiber. We suspect that such an asymmetry arises from the fact that the analyzed fiber shown in Fig. 3a changes its growth direction slightly. Therefore, an assumption of a fixed direction of lattice rotation axis is not fully justified. Nevertheless, the observations are consistent with the active role of dislocations having slip system {m}<a> (see Fig. S1 in SI).

The observed evidence that the crystal rotation axis is determined by its growth direction seems to be a determining factor for the spherulitic shape of region formed by the nucleation point and the subsequently grown crystalline areas. The fact that each growth direction has its own crystal rotation axis inhibits the formation of large 2D single crystals. Narrow fibers with individual points having their rotation axis (relative to the orientation of the nucleation point) in the vicinity of one crystal direction, are formed instead. From the previous analysis it also seems that these "allowed" orientations of the crystal rotation axis are determined by common slip systems, summarized for the Quartz crystal in Table S1 in SI.

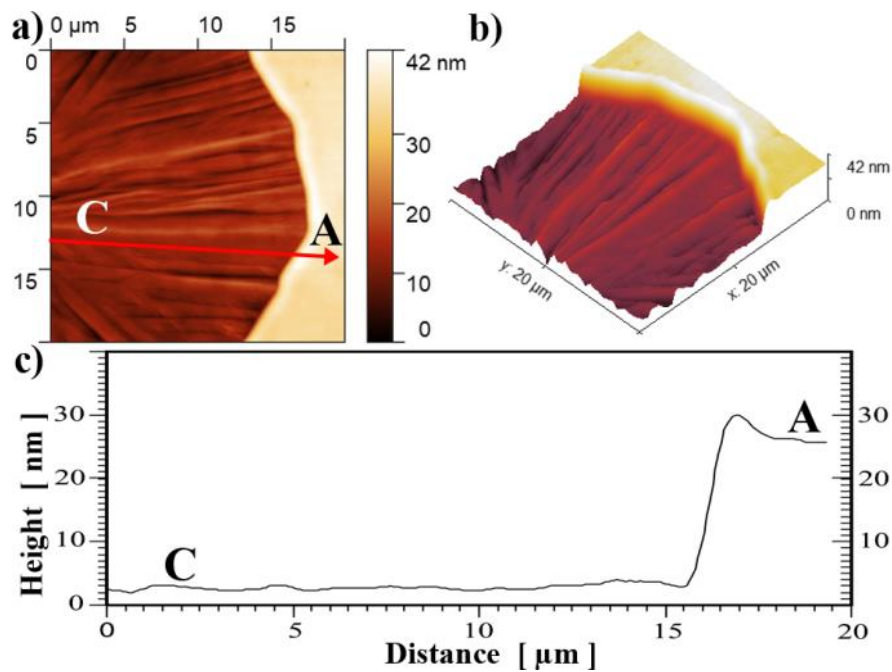

*Fig. 4: Crystalline/Amorphous (C/A) interface profile studied by AFM.*
  *a) Top view of spherulite/amorphous coating interface.*
  *b) The same area shown in 3D view.*
  *c) Height profile along the arrow shown in a). Height averaged over 9 lines close and parallel to the arrow.*

The origin of the trans-rotational crystal lattice formation analyzed above is also established using additional characterization techniques. Figure 4 shows the AFM observation of an area of the interface between the spherulitic crystal and its amorphous surrounding. A substantial sharp drop in height (23 nm) from the amorphous to the crystalline area is clearly observed. Moreover, a small (4 nm) glassy hill at the crystallization front is always present. Knowing the thickness of the deposited glassy film (~120 nm), and assuming mass conservation, we estimate that the density increase during crystallization is about 16.5 %. This value is similar to the difference in density between amorphous $GeO_2$ and its Quartz-type crystal structure (~18 %)[22]. The whole height change, between glassy and Quartz state, happens over a distance of about 600 nm (the change of profile in Fig. 4c does not appear so sharp due to the averaging effect of a few profile lines that are not perpendicular to the sharp interface). Therefore the slope of the surface profile along the interface is only ~ 3º over a distance of about 600 nm.

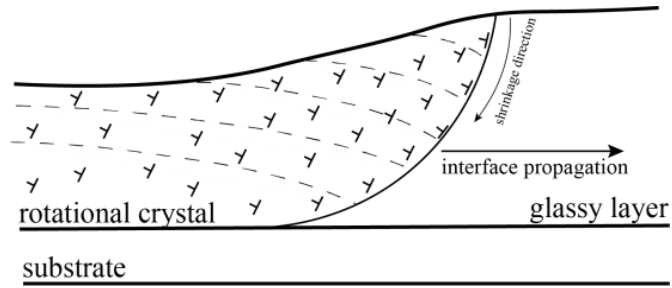

*Fig. 5: Schematic of the dislocation model of rotating crystal fiber growth from an amorphous film to explain the stress accommodation that takes place at the interface between low-density glass and high-density crystalline α-Quartz, based on ideas from[8,9]. Dashed lines represent the bent lattice planes.*

Using a slight modification of the models[8,9], we propose the following explanation, sketched in Fig. 6. We will assume that the crystallization front near the surface is ahead of the front in the vicinity of substrate and that this front propagates as a whole in a direction parallel to the film/substrate interface. During crystallization, substantial densification by shrinking along the glass/crystal interface must happen. For a thin film, the increase in density particularly occurs perpendicular to the free surface, because only in this direction the shape change is unconstrained[8]. From the observed 4 nm hill formed at the crystallization front, it seems that the rate of height shrinking along the crystallization front by diffusion is slower than the propagation rate of the glass/crystal interface. The formation of one type of dislocations on the crystalline side of this interface is anticipated by using an analogy to hetero-epitaxial crystal growth with misfit dislocations appearing periodically to compensate for the lattice parameter mismatch[9]. An additional half-plane appears on the crystal side, due to a smaller distance between atoms (higher density) of the crystalline state. These unpaired dislocations, initially appearing at the interface, remain in the growing crystal volume. They act as GNDs, generating a continual lattice bending, responsible for the spherulite structure. This is marked in Fig. 5 by a set of curved lattice planes, initially parallel to the surface (left side of Fig. 5), and gradually being bent towards the substrate during crystal growth.

Combining our observations with those of Savytskii and co-authors, we see that for this type of crystalline film growth, a crystalline substrate does not play a substantial role. Nuclei could be epitaxial with the crystalline substrate, but as the fibers start to grow and the crystallization front forms a shape similar to Fig. 5, the epitaxy between substrate and fiber disappears due to fiber lattice rotation. However, as we observed in Fig. 1, sometimes the nucleation of spherulite centers was not epitaxial.

A simple estimation of the maximum strain value in the crystal fiber is possible from Fig. 2a, by assuming an elastic cylindrical bending of the whole fiber. The maximum strain $\varepsilon$ of the surface film is estimated according to[7]:

$$\varepsilon = \frac{t}{2}\, grad(\varphi)\,, \tag{1}$$

where $t \approx 120\ nm$ is the thickness of the crystalline Quartz film, and $grad(\varphi)$ is the average measured gradient of monotonously increasing misorientation angle expressed in radians/m.

The resulting strain of 0.07% is relatively small. It is important to note that crystal growth occurred at relatively high temperatures, where the elastic limit is much smaller than the room temperature value.

The linear increase of the misorientation angle along the fiber is clear evidence that the same type of lattice deformation mechanism is active, leaving a constant density of GNDs behind the crystal growth front[23], responsible for the crystal lattice bending. The EBSD observation of these spherulitic fibers allows simultaneous determination of fiber nucleation area (orientation of non-deformed crystal) and the crystal rotation axis between a nucleation point and all other fiber points that are exposed to deformation. This generates a unique situation in which the detailed deformation mechanism analysis can be performed. We then use the well-established fact from dislocation theory that the crystal rotation axis between two parts of a crystal deformed by one type of dislocations is linked to their slip system, i.e. dislocation slip plane normal (SPN) and slip direction (SD), being the crystal rotation axis orthogonal to both of them. The SPN and SD are quite well categorized for Quartz crystals[24,25], as Table S1 in SI summarizes. Once the type of dislocation slip is known, the GND density can be estimated from the local crystal curvature, measured via misorientation angle gradient[26]. Therefore, an EBSD measurement allows to determine the type of the GNDs responsible for crystal rotation (from the main orientation of crystal rotation axis), as well as an estimate of their local densities (from the local crystal bending given by point-to-origin misorientation angle slope).

Let us estimate the GNDs density inside the fiber analyzed in Fig. 3. An average rotation angle gradient of 0.62 °/μm has been measured. Clear clustering of lattice rotation axes around the normal to basal plane (0001) is evident, which in a Quartz crystal corresponds to so-called {m}<±a> prismatic slip defined by SPN = {10$\bar{1}$0} and SD = <1$\bar{2}$10>[24]. A relatively simple formula can be used to roughly estimate the local density of GNDs if one assumes a homogeneous distribution[23,27]:

$$\rho_{hom} = \frac{\theta_{tot}}{|b|\,\Delta x}, \qquad (2)$$

where $\theta_{tot}$ is the lattice bending angle over distance $\Delta x$, realized by means of dislocations with Burger's vector *b*. Substituting our experimentally observed values: $\theta_{tot}$ = 0.422, $\Delta x$ = 39 µm and |b| being the size of the crystal axis a = 4.985 Å, we obtain an average GND density inside the analyzed fiber of 2.2 x$10^{13}$ m$^{-2}$. Reported dislocation densities in natural Quartz vary from low concentrations of about $10^9$ m$^{-2}$ in crystals, that have grown slowly from aqueous solutions in veins and cavities, to high density values of $10^{14}$ m$^{-2}$ observed in deformed rocks of low metamorphic grades[28]. Such a large dislocation density ($10^{14}$ m$^{-2}$) has also been observed after low temperature (580 ºC) creep deformation of Quartz crystal[29]. Dislocations in Quartz crystals have been known to cause problems in the fabrication of resonators by the formation of etch channels. It is also suspected that dislocations contribute to acceleration sensitivity, thermal hysteresis, and possibly aging of these devices[30]. In any case, a large body of literature on strain and deformation via dislocations in natural and cultured Quartz studied using EBSD has been published[5,24,25,31]. The presence of dislocations of the determined relatively high density (2.2 x$10^{13}$ m$^{-2}$) inside primary fibers could lead to an acceleration of non-crystallographic branching, such as crystal inhomogeneity, dislocations, and stress underlie non-crystallographic branching[3]. This mechanism is responsible for the formation of secondary lateral fibers, which by their subsequent growth, fill an initially empty amorphous space between primary fibers and thus finally form a 2D spherulitic shape.

Finally, we demonstrate that the same mechanism for spherulitic crystal growth takes place on a pure amorphous SiO$_2$ film which undergoes the following process: deposition by ALD and subsequently crystallization at a high temperature. A small amount of Sr has been pre-deposited on a Si substrate with the aim to promote crystallization[32]. The 2D spherulitic character of Quartz crystal growth is evident from the IPF map shown in Fig. 6a. The nucleation of the central area of all spherulitic colonies in this sample were epitaxial with the Si(100) substrate (see crystal orientation sketched in Fig. 6b). Yet the growth of individual fibers is again associated with a continuous crystal rotation, as shown by the pole figures in Fig. 6b). The primary fiber growing from the centrum in south-east direction has been selected for further crystal lattice misorientation analysis. The misorientation angle size map is shown in Fig. 6c, with lattice rotation angle gradient of 0.58°/μm, over the whole fiber length (61 μm).

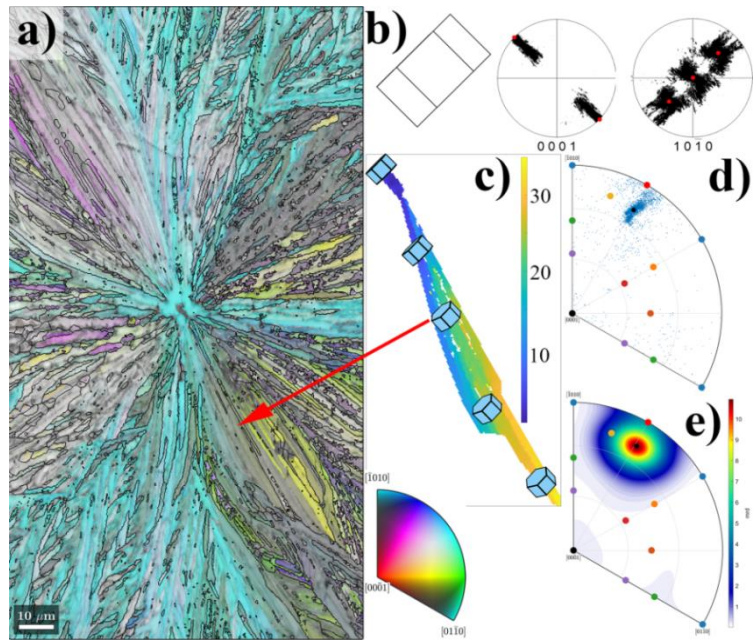

*Fig. 6: EBSD observations of spherulitic SiO$_2$ Quartz crystal:*
*a) Surface normal IPF map (color) combined with IQ map (grey scale) and grain boundary (>3°) black lines;*
*b) Quartz lattice orientation at the center of the spherulite, and (0001) and (10$\bar{1}$0) pole figure of the whole spherulite. The orientation of the spherulite center is marked by red points in the accompanying pole figures;*
*c) Misorientation angle size map (degrees) of one primary fiber growing in a south-east direction (of image in a)). The local lattice orientation is shown in a few points;*
*d) Rotation axis directions calculated between the reference point and all other fiber points. Mean axis direction is marked by a small black point.*
*e) Rotation axes distribution calculated from data shown in d).*
*The rotation axes of all common slip systems from Fig. 1c and Table 1 are also marked inside the IPFs.*

The insets of local lattice orientations in Fig. 6c) again suggest that the lattice rotation axis is almost perpendicular to the direction of fiber growth, as well as to the <c> axis of the lattice. The results of the exact calculations of these directions are summarized in Fig. 6d-e. Individual rotation axes as well as calculated distribution density are shown. It is important to note that neither the mean direction, nor the direction of maximum of their distribution is in the vicinity of a crystal rotation axis associated with a single type of dislocation deformation mechanism, marked by color points. A possible explanation is that the simultaneous action of a few dislocation types is present, and that the resulting maximum/mean rotation direction is a combine effect of two types of dislocations with different characteristic slip system and rotation axis. Using MTEX (Matlab extension package) we are able to calculate a rotation axis resulting from two subsequent crystal rotations over two different crystal axes. The situation is described in following diagram:

$$O_0 \xrightarrow{\alpha @ [A]} O_1 \xrightarrow{\beta @ [B]} O_2 \ . \tag{3}$$

An initial crystal orientation $O_0$ is transformed by rotation around a crystal axis with direction [A] by an angle $\alpha$, resulting in a new crystal orientation $O_1$. Subsequently orientation $O_1$ is transformed by rotation around a new crystal axis [B] by an angle $\beta$, resulting in a final crystal orientation $O_2$. We are interested in the misorientation angle $\gamma$ and the crystal direction [C] required for a direct transformation from the initial to the final orientation:

$$O_0 \xrightarrow{\gamma @ [C]} O_2 \ . \tag{4}$$

In this way, it is quite straightforward to simulate these rotations and determine $\gamma$ and [C] for any combination of rotation axes and misorientation angles from Eq(3). We calculated directions [C] using two subsequent and very small ($\approx 0.01°$) rotation angles $\alpha$ and $\beta$ around a few combinations of different rotation axes [A] and [B], each of them corresponding to another Quartz crystal slip system. The results of such calculations are shown in Fig. 7, and can be summarized in the following way: For small rotation angles $\alpha$ and $\beta$, the resulting rotation angle $\gamma$ is also small (the same order of magnitude) and the rotation direction [C] is located in the Quartz crystal IPF on a line directly connecting these two characteristic rotation axes, as Fig. 7 shows. An interesting fact is that the direction [C] does not depend on the order of the two subsequent rotations, but only on the ratio between rotation angles $\alpha$ and $\beta$. If the ratio is 1, the direction [C] is localized in an equidistant-point between [A] and [B] (small red dots in the middle of connection lines). Any change of the ratio between angles $\alpha$ and $\beta$ results in a shift of the position of [C] axis direction towards the direction with the larger rotation angle.

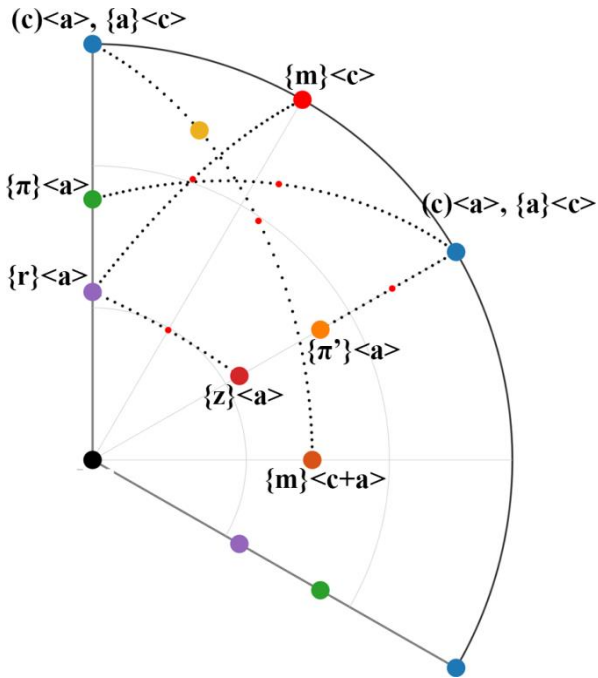

*Fig. 7: Quartz crystal IPF with colored directions of rotation axes characteristic for different slip systems according Table S1. Black dotted lines connecting two different directions correspond to calculated directions [C] that arise from a combination of rotations around these two characteristic directions. Small red dots mark the direction [C] that arises from two subsequent rotations with $\alpha = \beta$.*

On the basis of the results of the MTEX simulations mentioned above, and from the comparison of Fig. 7 and Fig. 6d)-e), we conclude that a presence of two types of dislocations could be responsible for small lattice rotations of the fiber analyzed in Fig. 6. Dislocations acting in slip system $\{\pi\}$<a> combined with one from (c)<a> or $\{a\}$<c> systems (see Fig. S1c in SI) are required to act simultaneously. This finally results in a Quartz lattice rotation around the crystallographic direction [$\bar{1}$ 0.507 0.493 0.404], determined to be the mean direction of all rotation axes acting in the fiber.

**Conclusions**

We use EBSD analysis to understand the spherulitic growth of crystalline Quartz from an amorphous thin film with unprecedented detail. The 2D spherulitic growth of thin Quartz crystalline is a consequence of rotational crystal growth. After nucleation, a unitary set of dislocations are generated at the glass/crystal interface to compensate for the substantial reduction of specific volume during crystallization. These dislocations form a cloud of GNDs with densities around $2.2 \times 10^{13}$ m$^{-2}$ inside each fiber and bend its crystalline lattice plane parallel to surface with a curvature in the range $5 \cdot 10^3 - 25 \cdot 10^3$ m$^{-1}$.

Due to the GNDs, the growing crystal is forced to rotate around a few different crystal axes, whose orientations depend on the growth direction. This restriction, together with the anisotropy of Quartz crystal growth rate, results in the formation of few primary fibers. A relatively high density of dislocations promotes non-crystallographic branching: the mechanism of secondary fiber nucleation. Subsequent growth of these fibers contributes to the complete filling of the spherulitic area with crystallites possessing deviated orientations.

The crystallization at relatively high temperatures supports an activation of many dislocation slip systems, so far reported mostly in geological Quartz. Quite often, even simultaneous action of two close slip systems is detected, when the crystal growth process forces the lattice to rotate around a crystal axis that is not in the vicinity of a lattice rotation realized by one single slip system.

**References:**

1. Zhou, S. *et al.* Crystallization of GeO2 thin films into alpha-quartz: from spherulites to single crystals. *Submitted for publication, arXiv:2007.03916 [cond-mat]* (2021).

2. Goldenfeld, N. Theory of spherulitic crystallization. *II-VI Compd 1987, Proc of the Third Int Conf on II-VI Compd* **84**, 601–608 (1987).

3. Shtukenberg, A. G., Punin, Y. O., Gunn, E. & Kahr, B. Spherulites. *Chem. Rev.* **112**, 1805–1838 (2012).

4. Thomas, A. *et al.* Mimicking the Growth of a Pathologic Biomineral: Shape Development and Structures of Calcium Oxalate Dihydrate in the Presence of Polyacrylic Acid. *Chemistry – A European Journal* **18**, 4000–4009 (2012).

5. Kilian, R. & Heilbronner, R. Analysis of crystallographic preferred orientations of experimentally deformed Black Hills Quartzite. *Solid Earth* **8**, 1095–1117 (2017).

6. Muscarella, L. A. *et al.* Crystal Orientation and Grain Size: Do They Determine Optoelectronic Properties of MAPbI$_3$ Perovskite? *J. Phys. Chem. Lett.* **10**, 6010–6018 (2019).

7. Kolosov, V. Y. & Thölén, A. R. Transmission electron microscopy studies of the specific structure of crystals formed by phase transition in iron oxide amorphous films. *Acta Materialia* **48**, 1829–1840 (2000).

8. Kooi, B. J. & De Hosson, J. Th. M. On the crystallization of thin films composed of Sb3.6Te with Ge for rewritable data storage. *Journal of Applied Physics* **95**, 4714–4721 (2004).

9. Savytskii, D., Jain, H., Tamura, N. & Dierolf, V. Rotating lattice single crystal architecture on the surface of glass. *Sci Rep* **6**, 36449 (2016).

10. Wright, S. I., Nowell, M. M., De Kloe, R. & Chan, L. Orientation precision of electron backscatter diffraction measurements near grain boundaries. *Microscopy and Microanalysis* **20**, 852–863 (2014).
**References:**

1. Zhou, S. *et al.* Crystallization of GeO2 thin films into alpha-quartz: from spherulites to single crystals. *Submitted for publication, arXiv:2007.03916 [cond-mat]* (2021).

2. Goldenfeld, N. Theory of spherulitic crystallization. *II-VI Compd 1987, Proc of the Third Int Conf on II-VI Compd* **84**, 601–608 (1987).

3. Shtukenberg, A. G., Punin, Y. O., Gunn, E. & Kahr, B. Spherulites. *Chem. Rev.* **112**, 1805–1838 (2012).

4. Thomas, A. *et al.* Mimicking the Growth of a Pathologic Biomineral: Shape Development and Structures of Calcium Oxalate Dihydrate in the Presence of Polyacrylic Acid. *Chemistry – A European Journal* **18**, 4000–4009 (2012).

5. Kilian, R. & Heilbronner, R. Analysis of crystallographic preferred orientations of experimentally deformed Black Hills Quartzite. *Solid Earth* **8**, 1095–1117 (2017).

6. Muscarella, L. A. *et al.* Crystal Orientation and Grain Size: Do They Determine Optoelectronic Properties of MAPbI$_3$ Perovskite? *J. Phys. Chem. Lett.* **10**, 6010–6018 (2019).

7. Kolosov, V. Y. & Thölén, A. R. Transmission electron microscopy studies of the specific structure of crystals formed by phase transition in iron oxide amorphous films. *Acta Materialia* **48**, 1829–1840 (2000).

8. Kooi, B. J. & De Hosson, J. Th. M. On the crystallization of thin films composed of Sb3.6Te with Ge for rewritable data storage. *Journal of Applied Physics* **95**, 4714–4721 (2004).

9. Savytskii, D., Jain, H., Tamura, N. & Dierolf, V. Rotating lattice single crystal architecture on the surface of glass. *Sci Rep* **6**, 36449 (2016).

10. Wright, S. I., Nowell, M. M., De Kloe, R. & Chan, L. Orientation precision of electron backscatter diffraction measurements near grain boundaries. *Microscopy and Microanalysis* **20**, 852–863 (2014).
24

**Acknowledgement**

The authors acknowledge financial support from NWO's TOP- 260 PUNT Grant No. 718.016002. Well organized support of MTEX package on [https://github.com/mtex-toolbox/mtex/discussions](https://github.com/mtex-toolbox/mtex/discussions) is also acknowledged.

**Supplementary Information**

Please contact authors to access the supplementary information (SI).